# Incommensurate spiral order from double exchange interactions

Maria Azhar and Maxim Mostovoy
*Zernike Institute for Advanced Materials, University of Groningen,*
*Nijenborgh 4, 9747 AG Groningen, The Netherlands*
(Dated: January 12, 2017)

The double exchange model describing interactions of itinerant electrons with localized spins is usually used to explain ferromagnetism in metals. We show that for a variety of crystal lattices of different dimensionalities and for a wide range of model parameters the ferromagnetic state is unstable against a non-collinear spiral magnetic order. We revisit the phase diagram of the double exchange model on a triangular lattice and show in a large part of the diagram the incommensurate spiral state has a lower energy than the previously discussed commensurate states. These results indicate that double exchange systems are inherently frustrated and can host unconventional spin orders.

*Introduction*: Non-collinear spin orders are inextricably related to inversion symmetry breaking in crystals and give rise to unconventional physical phenomena, such as magnetically-induced ferroelectricity and electric excitation of magnons in spiral magnets [1, 2]. Non-coplanar spin structures of skyrmions induce effective electromagnetic fields resulting in topological electron and magnon Hall effects [3]. Skyrmion dynamics induced by applied electric currents can be used in high-density magnetic memory devices [4]. Non-collinear magnetic orders are, however, relatively rare and it is of great interest to find new materials showing such states.

There are several well-understood microscopic mechanisms for non-collinear spin ordering. One of them is the relativistic Dzyaloshinskii-Moriya interaction which stabilizes spiral and skyrmion crystal states in chiral magnets [5, 6]. Non-collinear magnetism in Mott insulators is often a result of competing ferromagnetic (FM) and antiferromagnetic (AFM) Heisenberg interactions between spins, while in magnets with both itinerant and localized electrons it can originate from the RKKY interaction [7–9] closely related to Fermi surface instability.

Here we focus on the double exchange (DE), which was originally invoked to explain ferromagnetism in doped manganites [10, 11]. The DE model, also known as the ferromagnetic Kondo lattice model, describes a lattice of classical spins interacting with the conduction electrons through the Hund's rule coupling which aligns the spins of the conduction and localized electrons occupying the same lattice site. If the spins on neighboring sites are not parallel, the effective electron hopping amplitude decreases, which increases kinetic energy of the conduction electrons. In this way conduction electrons provide an effective FM interaction between the lattice spins.

This argument, however, cannot hold for all values of the model parameters, which is clear already from the fact that in the limit of small Hund's rule coupling constant, $J$, the model yields the RKKY interactions that can be FM or AFM, depending on the distance between the spins, which frustrates the uniform FM order and can lead to glass-like states [12, 13]. In the opposite limit of strong $J$, the electron hopping for the half-filled conduction band can be treated as a perturbation. It leads to an AFM exchange between spins, as in the Hubbard model, with $J$ playing the role of the on-site Coulomb repulsion $U$ [14]. Numerical and analytical studies showed that DE interactions can stabilize incommensurate [15] and various commensurate antiferromagnetic phases as well as a non-coplanar 'flux state' [14, 16–21]. The commensurate states are found close to particular electron filling fractions and the mechanism for their stabilization is the Fermi surface instability, i.e. opening of a (pseudo)gap in the spectrum of conduction electrons.

The Fermi surface instability is not the only way in which DE interactions can stabilize non-uniform magnetic orders. It was previously suggested that the long-period spiral state in the cubic perovskite $SrFeO_3$, which according to an x-ray photoemission study has a negative charge-transfer energy [22], can result from the coupling between localized spins formed by the Fe $d^5$ states and itinerant oxygen holes [23]. The spiral ordering affects electrons in the whole Fermi sea and does not require a nested Fermi surface. The transition between the collinear FM and non-collinear spiral states was also found in a two-dimensional model with a parabolic electron band [24, 25]. It occurs at a critical electron concentration, at which the electron chemical potential touches the bottom of the empty minority band. In the one-dimensional DE model the spiral state has a lower energy than the FM state for all electron concentrations, below a critical value of $J$ [26, 27].

In this Letter we show that the instability of the FM state towards the spiral ordering is a general property of the DE model: for a large variety of crystal lattices and a wide range of model parameters the incommensurate spiral state has a lower energy than the FM state. For Bravais lattices we give a simple analytical expression for the spin stiffness of the FM state that vanishes at the transition to the spiral state (non-Bravais lattices are discussed in the Supplemental Material [28]). We also show that the phase diagram of the DE model on a triangular lattice changes drastically when the spiral state is included: this state is lower in energy than the previously considered states and it fills the regions previously associated with phase separation.

*Instablity of the FM state*: The Hamiltonian of the DE model is the sum of the kinetic energy of the conduction electrons and the Hund's rule coupling:

$$H_{DE} = -\sum_{ij} t_{ij}\psi_i^\dagger \psi_j - J\sum_i \psi_i^\dagger \boldsymbol{\sigma}\psi_i \cdot \boldsymbol{S}_i, \quad (1)$$

Here, the operator $\psi_i = (\psi_{i\uparrow}, \psi_{i\downarrow})^T$ annihilates electron at the lattice site $i$ and $t_{ij}$ is the hopping amplitude: $t_{ij} = t$ for pairs of nearest-neighbor sites and is zero otherwise; $J$ is the strength of the Hund's rule coupling between the classical spin $\boldsymbol{S}_i$ of unit length and the conduction electron spin $\frac{1}{2}\psi_i^\dagger \boldsymbol{\sigma}\psi_i$ on the same site, where $\boldsymbol{\sigma} = (\sigma^x, \sigma^y, \sigma^z)$ is the vector composed of the three Pauli matrices.

To study the transition between the FM and spiral states, we calculate the energy of the spiral state with a wavevector $\boldsymbol{Q}$, in which spins rotate around the $x$-axis: $\boldsymbol{S}_i = (0, \sin\theta_i, \cos\theta_i)$, where $\theta_i = \boldsymbol{Q}\cdot\boldsymbol{x}_i$, $\boldsymbol{x}_i$ being the coordinate of the site $i$. We perform the transformation to the co-rotating spin frame, $\psi_i = e^{-i\frac{\sigma_x}{2}\theta_i}\psi_i'$, in which the spin vector is parallel to the $z$-axis at all lattice sites. In this frame the Hund's rule coupling has the same form as in the FM state, $-J\sum_i \psi_i'^\dagger \sigma_z \psi_i'$, while the kinetic energy, $T$, becomes $\boldsymbol{Q}$-dependent. For small spiral wave vectors, $T$ can be expanded in powers of $Q$:

$$\begin{aligned}T &\approx T^{(0)} + T^{(1)} + T^{(2)} \\ &= -\sum_{ij} t_{ij}\psi_i'^\dagger \left[1 + i\frac{\sigma_x}{2}(\theta_i - \theta_j) - \frac{1}{8}(\theta_i - \theta_j)^2\right]\psi_j'.\end{aligned} \quad (2)$$

To second order in $Q$ the difference between energies of the spiral and FM states is

$$\Delta E = -\sum_\nu \frac{|\langle \nu | T^{(1)} | 0\rangle|^2}{E_\nu - E_0} + \langle 0 | T^{(2)} | 0\rangle, \quad (3)$$

where $\nu$ labels excited electron states with one flipped spin. The first (negative) term in Eq. (3) results from the mixing of the occupied spin-up electron states with the unoccupied spin-down states, which lowers the kinetic energy [23]. The second term accounts for the increase in kinetic energy due to the band narrowing in the noncollinear spiral state. When the positive and negative terms are equal, the spin stiffness vanishes signaling the instability of the FM state. For Bravais lattices Eq.(3) can be written in the form [28]:

$$\Delta E = \frac{Q^2}{8d}\left[\frac{1}{2J}\int_{\mu-J}^{\mu+J} d\mu' E_0(\mu') \right. \\ \left. -\frac{1}{2}(E_0(\mu+J) + E_0(\mu-J))\right], \quad (4)$$

where $d$ is the lattice dimensionality, $\mu$ is the chemical potential, and $E_0$ is the total energy of the free electron state. If the function $E_0(\mu)$ is convex in the energy interval $[\mu-J, \mu+J]$, the Hermite-Hadamard inequality implies that $\Delta E < 0$, corresponding to instability of the FM state.

Equation (4) shows that the stabilization of the spiral state comes from a wide energy interval rather than from a narrow vicinity of the Fermi surface, as in the case of the spin density wave type of instability. This explains why the spiral state can exist in a wide range of the filling fraction, $x$ (the number of electrons per site divided by 2), and $J/t$ ratio, as shown in Fig. 1, where we plot the instability lines, $\Delta E = 0$, for eight two- and three-dimensional lattices. These lines separate shaded regions, where the collinear FM order is unstable ($\Delta E < 0$) from the unshaded FM regions, where $\Delta E > 0$.

For all lattices the FM state is unstable around half filling ($x = 0.5$) for large $J/t$, in which case the majority band is close to being completely filled and the energy increase due to the reduction of the band width (the second term in Eq.(3)) is relatively small. At half filling, perturbation theory in $t$ yields an effective nearest-neighbor antiferromagnetic Heisenberg exchange interaction with the exchange constant, $\frac{2t^2}{J}$, favoring antiferromagnetic states [14].

Stability of the FM state at small electron concentrations depends on the lattice dimensionality: for $d = 1$ or 2, there is a critical value of $J_c(x)$, below which the FM state is unstable, while for $d = 3$, the FM state is stable for all $J$. In the limit $J \ll t$ Eq.(4) becomes

$$\Delta E = -\frac{(QJ)^2}{24d}\frac{d}{d\mu}(\nu(\mu)\mu), \quad (5)$$

where $\nu(\mu)$ is the density of states for free electrons. For $x \ll 1$, $\nu \propto (\mu - \varepsilon_{min})^{d/2-1}$, where $\varepsilon_{min} = -zt < 0$ is the bottom of the conduction band, $z$ being the number of nearest-neighbor sites. Equation (5) then gives $\Delta E \propto (\varepsilon_{min} - \frac{d}{2}\mu)(\mu - \varepsilon_{min})^{d/2-1}$, valid for $\mu \gtrsim \varepsilon_{min}$, so that $\Delta E < 0$, for $d = 1, 2$, and $\Delta E > 0$, for $d = 3$.

The electron-hole symmetry of the DE model on bipartite lattices (square, honeycomb, cubic and bcc) makes the instability lines symmetric under $x \to 1 - x$. For non-bipartite lattices, the behaviors in the low-electron density and low-hole density limits can be markedly different. Thus the FM state on a triangular lattice is stable at low hole densities (see Fig. 1(b)). The behavior of this model for $x > 3/4$, corresponding to $\mu > 2t$, is governed by the saddle point in the electron dispersion, which gives rise to a logarithmic singularity in the density of states at $\mu = 2t$, $\nu(\mu) \propto \ln\frac{W}{|\mu-2t|}$, with $W$ is of the order of the bandwidth [28]. As follows from Eq.(5), the sharp decrease of $\nu(\mu)$ for $\mu > 2t$, makes $\Delta E > 0$ (at least, for small $J$), which explains the stability of the FM state.

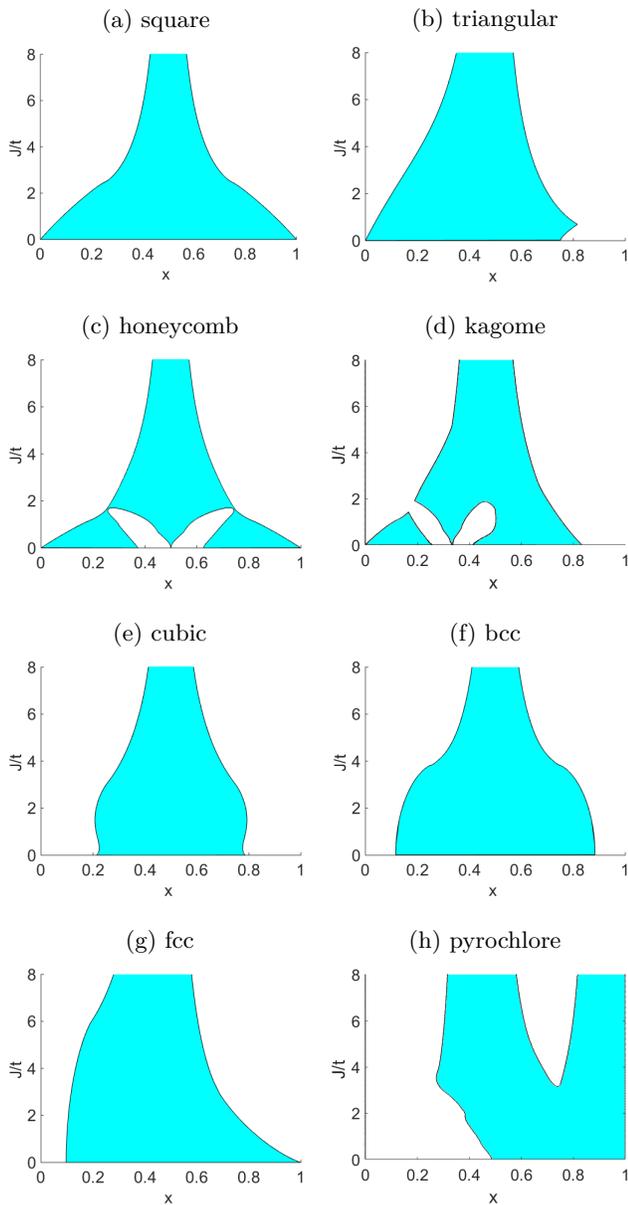

**Figure 1:** Instability of the FM state of the double exchange model on a (a) square (b) triangular (c) honeycomb (d) kagome (e) cubic (f) bcc (g) fcc and (h) pyrochlore lattice. At the instability lines (black solid lines) the spin stiffness of the FM state vanishes. These lines separate the shaded(white) regions where the spiral state has a lower(higher) energy than the FM state.

Another notable feature of the stability diagrams is the behavior near the Dirac points found for the honeycomb and Kagome lattices at the filling $\frac{1}{2}$ and $\frac{1}{3}$, respectively (see Figs. 1(c,d)). Due to the overlapping of the electron and hole Fermi surfaces, the first term in Eq.(3) logarithmically diverges, which favors the spiral state for $|\delta\mu| < J$, where $\delta\mu$ is the deviation of $\mu$ from its value at the Dirac point [28]. For $|\delta\mu| > J$, the divergent term is absent and the ground state is FM.

*Phase diagram for a triangular lattice*: Ferromagnetic and spiral states are not the only competing phases, as e.g. can be seen from the phase diagram of the DE model on a triangular lattice obtained in Ref. [14]. However, incommensurate spirals have not been considered in that as well as other numerical work done for relatively small sized lattices.

To include modulated magnetic states with a long period, we use the analytical expression for the electron dispersion in the spiral state with an arbitrary wave vector $\boldsymbol{Q}$,

$$\varepsilon_{\boldsymbol{k},\pm} = -2t \left( \sum_{\lambda=1}^{3} \cos(\boldsymbol{k}\cdot\boldsymbol{e}_\lambda) \cos\frac{\boldsymbol{Q}\cdot\boldsymbol{e}_\lambda}{2} \right)$$

$$\pm 2t \left[ \left( \sum_{\lambda=1}^{3} \sin(\boldsymbol{k}\cdot\boldsymbol{e}_\lambda) \sin\frac{\boldsymbol{Q}\cdot\boldsymbol{e}_\lambda}{2} \right)^2 + \left(\frac{J}{2t}\right)^2 \right]^{1/2}.$$

(6)

For a triangular lattice, $\boldsymbol{e}_1 = \boldsymbol{a}$, $\boldsymbol{e}_2 = \boldsymbol{b}$ and $\boldsymbol{e}_3 = -\boldsymbol{a}-\boldsymbol{b}$, $\boldsymbol{a}$ and $\boldsymbol{b}$ being the basis vectors of the hexagonal unit cell. The band structure of the spiral state consists of two bands, denoted by the ± symbol, resulting from mixing of the spin-up and spin-down states with the same wave vector $\boldsymbol{k}$. The energy splitting between the bands, and the predominant occupation of the lower-energy band stabilize the spiral state without opening a gap in the electron spectrum.

Using Eq.(6) we calculated the total energy of the spiral state and found the optimal $\boldsymbol{Q}$ as a function of $x$ and $J$. In addition, we analytically calculated the electron spectrum for all previously considered states of this DE model [14], which allowed us to compare energies of different states in the thermodynamic limit. The resulting phase diagram is shown in Fig. 2a. Different magnetic states are color coded and labeled as in Ref. [14].

The spiral state occupies a large part of the phase diagram and many states disappeared from the diagram after the spiral state was included. Moreover, the commensurate (3a) state with the 120° spin ordering which is stable near the half-filling, and the stripe antiferromagnetic state (2a) are special cases of the spiral state. All other states (excluding the uniform FM state) are concentrated in narrow regions near the three filling fractions, 0.22, 0.3 and 0.75, at which these small-period commensurate spin orders open a (pseudo)gap at the Fermi energy. Fig-ure 2a clearly shows that the mechanism stabilizing the spiral state is more general than the spin-density-wave instability that requires a nested Fermi surface.

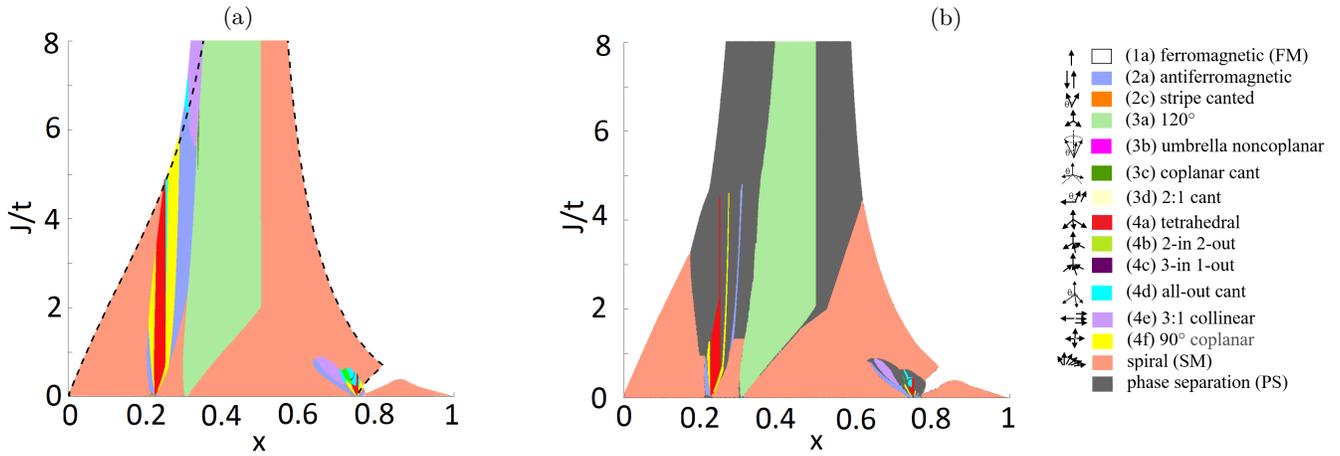

**Figure 2:** (a) Phase diagram of the double exchange model on the triangular lattice, obtained by comparing energies of different magnetically ordered states (see figure legend). The black dashed line is the boundary, obtained by perturbation theory in $\mathbf{Q}$, marks a continuous transition from a spiral state to a ferromagnetic state. (b) Phase diagram of the double exchange model on the triangular lattice showing the phase separation regions. The colored areas correspond to homogeneous magnetically ordered states minimizing the Gibbs energy. The diagrams have been obtained using a grid of $480 \times 480$ points in the Brillouin zone.

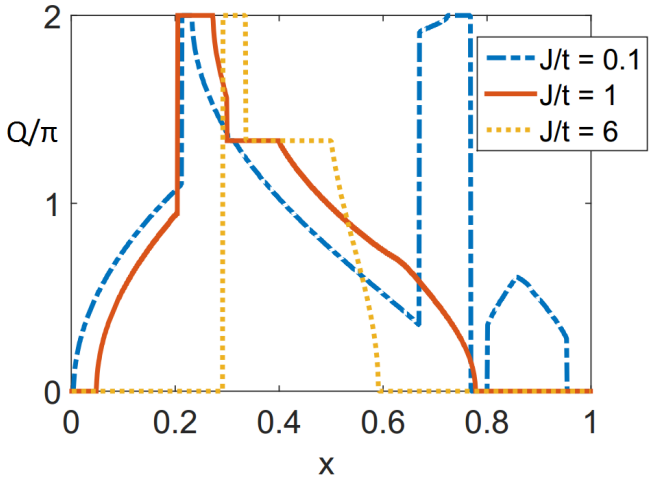

**Figure 3:** Magnitude of the wavevector $\mathbf{Q}$ of the spiral in units of $\pi$ versus the filling fraction $x$ at $J/t = 0.1$, 1 and 6 for the triangular lattice.

The energetically favored direction of the spiral wavevector $\mathbf{Q}$ is along $\pm \mathbf{e}_\lambda$, $\lambda = 1, 2, 3$. Figure 3 shows the magnitude of the spiral wave vector, $Q$, vs $x$ plotted for three $J/t$ ratios. For $J/t \sim 1$, $Q$ varies smoothly with $x$, except for the two plateaux at $Q = 2\pi$ and $Q = \frac{4\pi}{3}$ (the lattice constant is set to 1) corresponding to the commensurate states (2a) and (3a), respectively. For $J/t \gg 1$, the incommensurate spiral region shrinks. For $J/t \ll 1$ (the RKKY regime), the $x$-dependence of $Q$ is sensitive to the shape of the Fermi Surface [28]. While Fig. 1b shows that the FM state is stable for $x > 0.75$ and small $J/t$, in the phase diagram Fig. 2a this region is occupied by a spiral state. Figure 3 explains this apparent paradox. The spiral in this region is stabilized by the Fermi surface instability and the transition between the FM and spiral states is discontinuous: $Q$ jumps from 0 to a finite value.

So far, we discussed spatially homogeneous phases. To find the regions in the phase diagram where phase separation (PS) occurs, we minimize the Gibbs free energy, $G = E - \mu N$, where $N$ is the number of electrons, as a function of $\mu$. In general, phase transitions between different states are accompanied by a jump in $x$, and in the PS regions (black color) phases with the same $\mu$ and different $x$ coexist. Figure 2b shows that the spiral is stable against the phase separation in a large part of the phase diagram. Our phase diagram looks very different from that obtained in Ref. [14] where the incommensurate spiral state was not considered. Even when an additional antiferromagnetic exchange interaction between the local spins is switched on, the spiral state still remains the ground state in a large part of the diagram [28].

*Conclusions*: Our results show that non-collinear magnetism must be commonly present in double exchange systems. As the charge carrier density increases, the FM state eventually looses spin stiffness and undergoes a transition into an incommensurate spiral state. The driving force behind this instability is an additional splitting between the majority and minority electron states in the spiral phase. The splitting affects electrons with the same momentum and does not necessarily open a gap, which is why this mechanism works in metals and is very robust: we have found spiral states in a wide range of band fillings and $J/t$ ratio for lattices of different types and dimensionalities. Our results also apply to double exchange systems with orbital degeneracy [30], since the non-Bravais honeycomb, kagome and pyrocholore lattices can be considered as block lattices with several electron orbitals per block. The instability of the FM state can lead to more complex non-collinear and non-coplanar magnetic orders: an applied magnetic field, magnetic anisotropies and thermal fluctuations can stabilize multiply periodic states, such as the skyrmion crystal [31–33], which may explain the complexity of the phase diagram of the itinerant cubic magnet, SrFeO$_3$ [34].

---

# Incommensurate spiral order from double exchange interactions - Supplemental Material

Maria Azhar and Maxim Mostovoy
*Zernike Institute for Advanced Materials, University of Groningen,*
*Nijenborgh 4, 9747 AG Groningen, The Netherlands*

(Dated: January 12, 2017)

### Appendix A: Bravais lattices

In this section we derive Eq.(4). Equation (3) can be written as,

$$\Delta E = -\frac{1}{8J} Q^a Q^b \sum_{\mathbf{k}} (n_{\mathbf{k}\uparrow} - n_{\mathbf{k}\downarrow}) \frac{\partial \varepsilon_{\mathbf{k}}}{\partial k_a} \frac{\partial \varepsilon_{\mathbf{k}}}{\partial k_b} \\ + \frac{1}{8} Q^a Q^b \sum_{\mathbf{k}} (n_{\mathbf{k}\uparrow} + n_{\mathbf{k}\downarrow}) \frac{\partial^2 \varepsilon_{\mathbf{k}}}{\partial k_a \partial k_b}, \quad (A1)$$

where $n_{\mathbf{k}\uparrow}(n_{\mathbf{k}\downarrow})$ is the occupation number of the state with the wave vector $\mathbf{k}$ and spin up(down) in the co-rotating spin frame with the energy $\varepsilon_{\mathbf{k}\sigma} = \varepsilon_{\mathbf{k}} \mp J$, where

$$\varepsilon_{\mathbf{k}} = -2t \sum_{\lambda} \cos(\mathbf{k} \cdot \mathbf{e}_{\lambda}), \quad (A2)$$

$\mathbf{e}_{\lambda}$ being the vectors along the independent nearest-neighbor bonds for each lattice site, and the $-(+)$ sign corresponds to $\sigma = \uparrow (\downarrow)$. Hence, $n_{\mathbf{k}\sigma} = f(\varepsilon_{\mathbf{k}} - \mu \mp J)$, where $f(\xi) = \left(e^{-\beta\xi} + 1\right)^{-1}$ is the Fermi-Dirac distribution function.

Due to the rotational symmetry of all Bravais lattices considered in this Letter, the sums over $\mathbf{k}$ in Eq.(A1) are proportional to $\delta^{ab}$, so that

$$\Delta E = \\ \frac{Q^2}{4d} \Bigg[ -\frac{1}{2J} \sum_{\mathbf{k}} \Big( f(\varepsilon_{\mathbf{k}} - J - \mu) - f(\varepsilon_{\mathbf{k}} + J - \mu) \Big) \mathbf{v}_{\mathbf{k}}^2 \\ + \frac{1}{2} \sum_{\mathbf{k}} \Big( f(\varepsilon_{\mathbf{k}} - J - \mu) + f(\varepsilon_{\mathbf{k}} + J - \mu) \Big) \frac{\partial \mathbf{v}_{\mathbf{k}}}{\partial \mathbf{k}} \Bigg], \\ (A3)$$

where $d$ is the lattice dimensionality and $\mathbf{v}_{\mathbf{k}} = \frac{\partial \varepsilon_{\mathbf{k}}}{\partial \mathbf{k}}$ is the electron velocity.

The first term in the square brackets can be written as

$$\frac{1}{2J} \sum_{\mathbf{k}} \int_{\mu-J}^{\mu+J} d\mu' \frac{\partial f(\varepsilon_{\mathbf{k}} - \mu')}{\partial \varepsilon_{\mathbf{k}}} (\mathbf{v}_{\mathbf{k}} \cdot \mathbf{v}_{\mathbf{k}}) \\ = \frac{1}{2J} \sum_{\mathbf{k}} \int_{\mu-J}^{\mu+J} d\mu' \frac{\partial f(\varepsilon_{\mathbf{k}} - \mu')}{\partial \mathbf{k}} \cdot \mathbf{v}_{\mathbf{k}} \quad (A4) \\ = -\frac{1}{2J} \sum_{\mathbf{k}} \int_{\mu-J}^{\mu+J} d\mu' f(\varepsilon_{\mathbf{k}} - \mu') \frac{\partial \mathbf{v}_{\mathbf{k}}}{\partial \mathbf{k}}.$$

From Eq.(A2) we find $\frac{\partial \mathbf{v}_{\mathbf{k}}}{\partial \mathbf{k}} = -\varepsilon_{\mathbf{k}}$, as the length of all bonds connecting nearest-neighbor sites equals 1. From Eqs.(A3) and (A4) we then obtain Eq.(4) in the main text, where

$$E_0(\mu) = 2 \int_{-\infty}^{+\infty} d\varepsilon \nu(\varepsilon) \varepsilon f(\varepsilon - \mu) = 2 \int_{-\infty}^{\mu} d\varepsilon \nu(\varepsilon) \varepsilon \quad (A5)$$

is the free electron energy at zero temperature, $\nu(\varepsilon)$ being the density of states per spin projection.

### Appendix B: Non-Bravais lattices

For non-Bravais lattices, the free electron Hamiltonian $\hat{h}_{\mathbf{k}}$ in the $\mathbf{k}$-space (the Fourier transform of the hopping term) is a $n \times n$ matrix, where $n$ is the number of lattice sites in the unit cell. Its spectrum consists of $n$ bands, labeled by $\lambda = 1, \ldots, n$:

$$\hat{h}_{\mathbf{k}} |\mathbf{k}\lambda\rangle = \varepsilon_{\mathbf{k}\lambda} |\mathbf{k}\lambda\rangle. \quad (B1)$$

The analogue of Eq. (A1) is

$$\Delta E = \frac{Q^2}{4d} \times \\ \Bigg[ \sum_{\mathbf{k}\lambda\lambda'} \frac{|\langle \mathbf{k}\lambda'| \hat{\mathbf{v}}_{\mathbf{k}} |\mathbf{k}\lambda\rangle|^2}{\varepsilon_{\mathbf{k}\lambda} - \varepsilon_{\mathbf{k}\lambda'} - 2J} \left( f(\varepsilon_{\mathbf{k}\lambda} - \mu - J) - f(\varepsilon_{\mathbf{k}\lambda'} - \mu + J) \right) \\ + \frac{1}{2} \sum_{\mathbf{k}\lambda} \langle \mathbf{k}\lambda| \frac{\partial \hat{\mathbf{v}}_{\mathbf{k}}}{\partial \mathbf{k}} |\mathbf{k}\lambda\rangle \left( f(\varepsilon_{\mathbf{k}\lambda} - \mu - J) + f(\varepsilon_{\mathbf{k}\lambda} - \mu + J) \right) \Bigg], \\ (B2)$$

where $\hat{\mathbf{v}}_{\mathbf{k}} = \frac{\partial \hat{h}_{\mathbf{k}}}{\partial \mathbf{k}}$ is the velocity operator.

The intraband contribution to the first term in Eq.(B2) ($\lambda' = \lambda$) can be re-written as (cf. Eq.(A4))

$$-\frac{1}{2J} \sum_{\mathbf{k}\lambda} \int_{\mu-J}^{\mu+J} d\mu' f(\varepsilon_{\mathbf{k}\lambda} - \mu') \frac{\partial \mathbf{v}_{\mathbf{k}\lambda}}{\partial \mathbf{k}}, \quad (B3)$$

where $\mathbf{v}_{\mathbf{k}\lambda} = \langle \mathbf{k}\lambda| \hat{\mathbf{v}}_{\mathbf{k}} |\mathbf{k}\lambda\rangle$ is the electron velocity in the band $\lambda$. However,

$$\frac{\partial \mathbf{v}_{\mathbf{k}\lambda}}{\partial \mathbf{k}} = \langle \mathbf{k}\lambda| \frac{\partial \hat{\mathbf{v}}_{\mathbf{k}}}{\partial \mathbf{k}} |\mathbf{k}\lambda\rangle + 2 \sum_{\lambda' \neq \lambda} \frac{|\langle \mathbf{k}\lambda'| \hat{\mathbf{v}}_{\mathbf{k}} |\mathbf{k}\lambda\rangle|^2}{\varepsilon_{\mathbf{k}\lambda} - \varepsilon_{\mathbf{k}\lambda'}} \quad (B4)$$

contains interband contributions.

Using $\frac{\partial \hat{\mathbf{v}}_{\mathbf{k}}}{\partial \mathbf{k}} = -\hat{h}_{\mathbf{k}}$, we can combine all intra-band contributions into an equation similar to Eq.(4) for a Bravais lattice

$$\Delta E_{intra} = \frac{Q^2}{8d} \sum_{\lambda} \left[ \frac{1}{2J} \int_{\mu-J}^{\mu+J} d\mu' E_{\lambda 0}(\mu') \right. \\ \left. - \frac{1}{2} (E_{\lambda 0}(\mu+J) + E_{\lambda 0}(\mu-J)) \right], \quad (B5)$$

where $E_{\lambda 0}(\mu)$ is the free electron energy for the band $\lambda$.

The contribution of the interband transitions can be cast into the form,

$$\Delta E_{inter} = \frac{Q^2}{8d} {\sum_{\mathbf{k}\lambda\lambda'}}' |\langle \mathbf{k}\lambda'| \hat{\mathbf{v}}_{\mathbf{k}} |\mathbf{k}\lambda\rangle|^2 \times \\ \left[ -\frac{1}{2J} \int_{\mu-J}^{\mu+J} d\mu' \frac{f(\varepsilon_{\mathbf{k}\lambda}-\mu')-f(\varepsilon_{\mathbf{k}\lambda'}-\mu')}{\varepsilon_{\mathbf{k}\lambda}-\varepsilon_{\mathbf{k}\lambda'}} \right. \\ \left. + \frac{f(\varepsilon_{\mathbf{k}\lambda}-\mu+J)-f(\varepsilon_{\mathbf{k}\lambda'}-\mu-J)}{\varepsilon_{\mathbf{k}\lambda}-\varepsilon_{\mathbf{k}\lambda'}+2J} \right], \quad (B6)$$

where $\sum'$ means that $\lambda' \neq \lambda$.

### Appendix C: The saddle point

Electron spectra for all two-dimensional lattices considered in this Letter have one or more saddle points. The filling fraction at which the chemical potential of free electrons equals the saddle-point energy is $\frac{1}{2}$ for the square lattice, $\frac{3}{4}$ for the triangular lattice, $\frac{3}{8}$ and $\frac{5}{8}$ for the honeycomb lattice and $\frac{1}{4}$ and $\frac{5}{12}$ fillings for the kagome lattice.

Near the saddle point the electron energy is given by

$$\varepsilon_{\mathbf{k}\lambda} \approx ak_1^2 - bk_2^2 + \varepsilon_0, \quad (C1)$$

where $a, b > 0$ and $k_1$ and $k_2$ are components of the electron wave vector along two mutually orthogonal directions. This dispersion leads to a divergent density of electron states

$$\nu_\lambda(\varepsilon) = \frac{S}{8\pi^2 \sqrt{ab}} \ln \frac{W}{|\varepsilon|}, \quad (C2)$$

where $S$ is the area of the lattice and $W$ is the energy cut-off of the order of the band width.

While for a Bravais lattice one can use Eq.(5) to discuss the stability of the FM state near the saddle point, for non-Bravais lattices the argument has to be modified because, in general, $\text{div}(\mathbf{v}_{\mathbf{k}\lambda}) = \frac{\partial \mathbf{v}_{\mathbf{k}\lambda}}{\partial \mathbf{k}} \neq -\varepsilon_{\mathbf{k}\lambda}$. In particular, Eq.(C1) gives $\frac{\partial \mathbf{v}_{\mathbf{k}\lambda}}{\partial \mathbf{k}} = 2(a-b)$, which, in general, is different from $-\varepsilon_0$.

Using Eq.(B2) the singular contribution to $\Delta E$ coming from the intraband terms can be written in the form

$$\Delta E_{sing} = \frac{Q^2}{8}(a-b) \sum_\lambda \left[ \frac{1}{2J} \int_{\mu-J}^{\mu+J} d\mu' N_{\lambda 0}(\mu') \right. \\ \left. - \frac{1}{2}(N_{\lambda 0}(\mu+J) + N_{\lambda 0}(\mu-J)) \right], \quad (C3)$$

where $N_{\lambda 0} = 2\int_{-\infty}^{\mu} d\varepsilon \nu_\lambda(\varepsilon)$ is the number of electrons in the band $\lambda$ and we replaced $\text{div}(\mathbf{v}_{\mathbf{k}\lambda})$ by its value at the saddle point, $2(a-b)$, (see Eq.(C1) to keep only the most singular term.

Expansion of Eq.(C3) at small $J$ gives,

$$\Delta E_{sing} = (a-b)\frac{(QJ)^2}{24}\frac{d\nu_\lambda}{d\mu} \propto \frac{b-a}{\mu-\varepsilon_0}. \quad (C4)$$

For $b > a$, the incommensurate spiral ordering occurs for $\mu < \varepsilon_0$ ($\Delta E_{sing} < 0$), while for $\mu > \varepsilon_0$ the FM state is stable. This is the behavior found for the triangular lattice at $\frac{3}{4}$ filling, for the honeycomb lattice at $\frac{3}{8}$ filling and for the kagome lattice at $\frac{1}{4}$ filling. On the other hand, for the saddle point at $\frac{5}{8}$ filling of the honeycomb lattice $a > b$, so that the FM state is stable for $\mu < \varepsilon_0$, while the spiral state appears for $\mu > \varepsilon_0$. This explains the transitions at small $J$ seen in Figure 1 of the main text, at the saddle points.

### Appendix D: The Dirac point

The Dirac point $\mathbf{k} = \mathbf{k}_D$ is the crossing point of two bands, denoted by $\pm$, with the energy

$$\varepsilon_{\mathbf{k}\pm} = \varepsilon_0 \pm v |\mathbf{k} - \mathbf{k}_D|. \quad (D1)$$

The singular contribution to $\Delta E$ comes in this case from the transitions between the bands given by Eq.(B6). The absolute value of the matrix element $|\langle \mathbf{k}+| \hat{\mathbf{v}}_{\mathbf{k}} |\mathbf{k}-\rangle| = v$ is energy independent and the singularity originates from the fact that at $\mu = \varepsilon_0$ the electron-like Fermi surface of the spin-up band coincides with the hole-like Fermi surface of the spin-down band.

From Eq.(B6) we obtain

$$\Delta E_{\text{sing}} = -\frac{JQ^2}{32\pi}\theta(J-|\mu-\varepsilon_0|)\ln\frac{J}{|\mu-\varepsilon_0|}, \quad (D2)$$

where $\theta(x)$ is the Heaviside step function. This negative singular term corresponds to instability of the FM state for $|\mu-\varepsilon_0| < J$, found for the honeycomb lattice at half filling and the kagome lattice at $x = \frac{1}{3}$ (see Figs. 1(c) and (d)). For $|\mu-\varepsilon_0| > J$ the singularity disappears and the FM ordering is stable for these lattices.

## Appendix E: Spiral wave vector for small $J/t$

For $\frac{J}{t} \ll 1$, the spiral state results from the spin-density-wave instability: the Hund's rule coupling to rotating spins provides a perturbation that mixes electron states at nested parts of the Fermi surface connected by the spiral wave-vector $\mathbf{Q}$. The wave-vectors that connect more than one pairs of nested parts of the Fermi surface within the Brillouin zone are energetically preferred. In the small $J$ limit, the $x$-dependence of $Q$ (see Fig. (1)) is sensitive to the shape of the Fermi surface and does not have the plateaux that appear at larger $J/t$, associated with the 120-degrees and stripe antiferromagnetic orders.

At special filling fractions, multi-Q magnetic ordering may be stabilised through the opening of a local gap if the Q-vectors simultaneously connect multiple pairs of nested parts of the Fermi surface, and these orders are lower than the spiral ordering at fourth order in $J/t$ at these filling fractions [14].

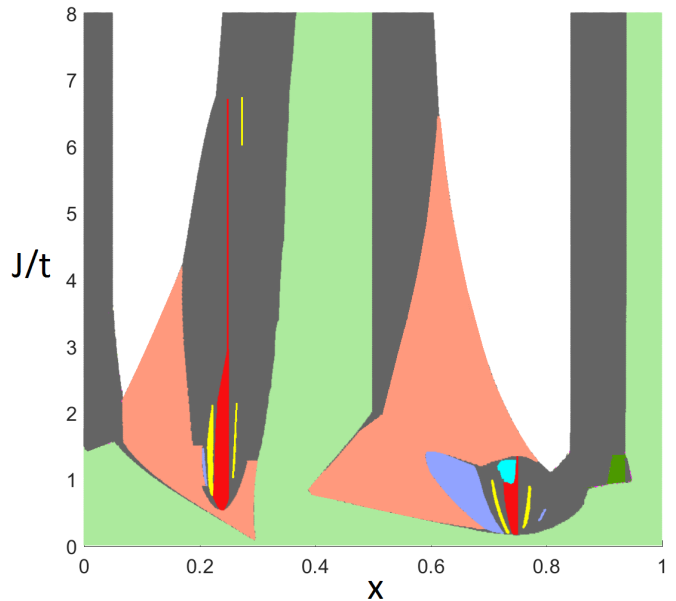

**Figure 2:** Phase diagram for the double exchange model on the triangular lattice in the presence of the antiferromagnetic exchange ($J_{AFM} = 0.01t$) between the local spins.

To calculate the regions of phase separation, we minimized the grand canonical potential for $10^4$ different values of $\mu$. A grid of $480 \times 480$ points in the Brillouin zone was used.

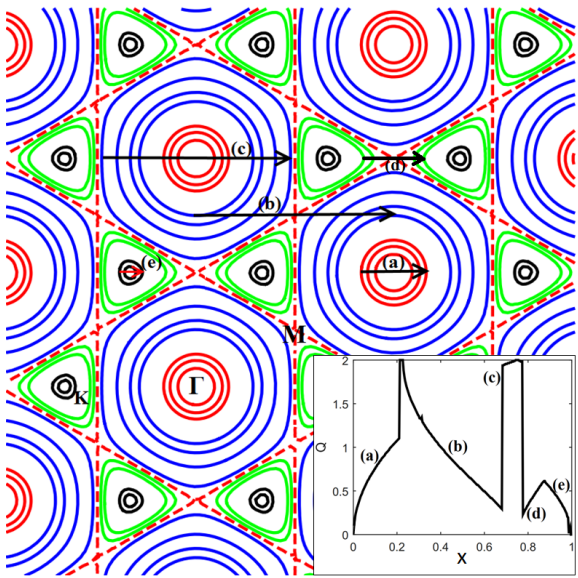

**Figure 1:** Dispersion of itinerant electrons on the triangular lattice. The red dashed line shows the Fermi surface at $\frac{3}{4}$ filling. The inset shows the $x$-dependence of $\mathbf{Q}$ for $J/t = 0.01$. Shown also are the wave vectors corresponding to the points labeled by a,b,…,e.

## Appendix F: Effect of weak antiferromagnetic interaction between spins

The antiferromagnetic nearest-neighbour Heisenberg exchange interaction betweed local spins, $H_{AFM} = J_{AFM} \sum_{<i,j>} \mathbf{S}_i \cdot \mathbf{S}_j$, favors the 120-degrees ordering. Still the spiral state occupies a large part of the phase diagram, as can be seen from the phase diagram in Fig. (2), where it replaces regions previously occupied by phase separation, ferromagnetism, and the collinear (2a) and (4e) magnetic orderings in the phase diagram of Ref. [29].